\documentclass[12pt]{article}

\usepackage{newtxtext,newtxmath}

\usepackage{graphicx}
\usepackage{gensymb}
\usepackage[letterpaper,margin=1in]{geometry}

\usepackage{siunitx}

\usepackage{multirow}

\renewenvironment{abstract}
	{\quotation}
	{\endquotation}

\date{}

\makeatletter
\renewcommand{\fnum@figure}{\textbf{Figure \thefigure}}
\renewcommand{\fnum@table}{\textbf{Table \thetable}}
\makeatother

\usepackage{scicite}

\usepackage{url}

\def\scititle{
	Mapping Temperature Using Transmission Kikuchi Diffraction
}
\title{\bfseries \boldmath \scititle}

\author{
	Yueyun~Chen$^{1,2}$,
	Xin~Yi~Ling$^{2}$,
	Jared~Lodico$^{1,2}$,
    Tristan~P.~O'Neill$^{1,2}$,\and
    B.~C.~Regan$^{1,2}$,
    Matthew~Mecklenburg$^{2,3\ast}$\and
	\small$^{1}$Department of Physics and Astronomy, University of California, Los Angeles, CA 90095, USA.\and
	\small$^{2}$California NanoSystems Institute (CNSI), University of California, Los Angeles, CA 90095, USA.\and
    \small$^{3}$Core Center of Excellence in Nano Imaging (CNI), University of Southern California, Los Angeles, \\
    \small CA 90009, USA. \and
	\small$^\ast$Corresponding author. Email: mmecklenburg@cnsi.ucla.edu
}

\begin{document} 

\maketitle

\begin{abstract} \bfseries \boldmath
Electronic devices are engineered at increasingly smaller length scales; new metrologies to understand nanoscale thermodynamics are needed. Temperature and pressure are fundamental thermodynamic quantities whose nanoscale measurement is challenging as physical contact inevitably perturbs the system. Here we demonstrate Kikuchi diffraction thermometry (KDTh), a non-contact scanning electron microscope (SEM) technique capable of mapping nanoscale temperatures and pressures. KDTh detects local volumetric lattice changes in crystalline samples by precisely fitting Kikuchi patterns. Temperature changes are deduced using the coefficient of thermal expansion (CTE). We map lattice parameters and temperatures on Joule-heated graphite by rastering a 5.5-nm electron probe across the sample. Our parameter precision is $\sim$0.01\% and our temperature sensitivity is 2.2~K/$\sqrt{\text{Hz}}$. KDTh offers advanced sensitivity by fitting the entire Kikuchi pattern, even beyond the precision measured in transmission electron microscopy. KDTh can operate in both transmission (transmission Kikuchi diffraction, TKD) and reflection (electron backscatter diffraction, EBSD) modes.
\end{abstract}


\noindent
To maximize the performance and durability of electronic devices with decreasing feature sizes, it is necessary to understand heat accumulation and dissipation \textit{in operando} at the microscopic level. A variety of metrologies, broadly classified into two categories, can determine the localized temperature: contact and non-contact thermometries. Contact thermometries require thermal contact between the sensor and the device under test (DUT). For example, in scanning thermal microscopy (SThM), a thermocouple probe is brought into contact with the DUT and a small amount of heat is transferred to induce a measurable voltage \cite{stopka_surface_1995}. These techniques are very sensitive; however, thermal contact disturbs the nanoscale system, and signal interpretation is complicated by the change in thermal contact during scanning \cite{menges_temperature_2016}. Non-contact thermometries collect temperature-dependent radiation such as infrared \cite{presotto_super-resolution_2024}, Raman \cite{reparaz_novel_2014}, and fluorescence \cite{kim_examination_2003}. Without a thermal contact that disturbs the sample, these non-contact thermometries are more repeatable, but their resolution is diffraction limited by the optical wavelength to about \SI{1}{\micro\metre}. Therefore, non-contact thermometry with sub-optical resolution is ideal. 

Electron microscopes have an atomic-level spatial resolution, making them a great platform for developing high-resolution non-contact thermometries. This microscopy is classified into three image modalities: transmission electron microscope (TEM), scanning transmission electron microscope (STEM), and scanning electron microscope (SEM). In recent years, thermometries have been developed for each. In TEMs, where the sample is illuminated by a parallel beam, the reciprocal lattice vectors \cite{cremons_direct_2016,fritsch_sub-kelvin_2022} and the Debye-Waller effect \cite{cremons_direct_2016} measure temperature with high sensitivity. However, the large parallel beam diameter, typically $>$100~nm, inhibits the temperature mapping capability. In STEMs and SEMs, a convergent electron probe is scanned across the sample, and information is acquired per point sequentially, allowing the temperature to be mapped in parallel. In STEMs, electron energy loss spectroscopy (EELS) can map temperature using an atomic-scale electron probe \cite{mecklenburg_nanoscale_2015,mecklenburg_visualizing_2021}. However, the temperature being mapped is well defined only at length scales larger than the mean free paths (MFP) of heat carriers (i.e. phonons and electrons), typically around a few nanometers near room temperature. Atomic-scale probes become delocalized when mapping thermal transport. In addition, TEMs and STEMs require extensive sample preparation for electron transparency and thermal gradient generation. Their field of view is also limited due to electron optics. SEMs require minimal sample preparation, have wide fields of view, and simple electrical connections for heat generation. In SEMs, electron backscattered diffraction (EBSD) has detected temperature-induced changes through the Debye-Waller effect and thermal diffuse scattering \cite{wu_novel_2012,gnabasik_fast_2025}. However, inelastic scattering is challenging to quantify and the spatial resolution of EBSD is limited to $\approx$100~nm by the interaction volume.

Transmission Kikuchi diffraction (TKD) collects forward-scattered electrons as the beam rasters across an electron-transparent sample in an SEM. TKD has a much smaller interaction volume compared to EBSD, thus offering sub-10~nm resolution \cite{shen_spatial_2019} and a higher pattern quality \cite{zhang_comparison_2025}. Here, we introduce Kikuchi diffraction thermometry (KDTh), a non-contact method for mapping temperature in an SEM, which is fast, precise, and accessible. Moreover, KDTh takes advantage of the pattern matching algorithm in the Kikuchi diffraction-based crystal orientation determination process to directly detect any temperature- or stress-induced lattice strain.

The network of Kikuchi lines scattered from a focused electron probe (few nanometers in size) provides information to the highest scattering angle, out past 1000 mrads (see Supplementary Information ``Collection Angles"). Diffraction information at such large angles is impossible to gather in TEMs or STEMs because of limitations by the Debye-Waller effect and bottom pole piece geometry. The atomic thermal and quantum motion lead to an uncertainty of $d_{\text{DW}} \sim$ 10~pm in the atomic position \cite{shevitski_dark-field_2013,khidirov_root-mean-square_2021}. These small variations produce correspondingly large momentum attenuation in coherent diffraction, which limits Bragg diffraction to $\lambda/d_{\text{DW}} \sim$ 200~mrads. For context, the TEM's bottom pole piece also limits the scattering to $\theta_{\text{max}} \approx (D/2)/f \approx$ (0.4~mm/2)/2~mm $\approx 100$~mrads, where $D$ is the bore diameter of the bottom pole piece and $f$ is the back focal length of the lower objective lens \cite{tsuno_design_1998}. Kikuchi-based indexing uses the total diffraction pattern area, so differences in scattering angles become more pronounced when their squares are compared. In SEMs, Kikuchi patterns are formed by electrons that undergo inelastic thermal diffuse scattering (TDS) followed by coherent Bragg diffraction \cite{liu_depth_2019}. This combination of inelastic and coherent scattering extends past the limit set by the Debye-Waller effect. In an SEM, which has no bottom pole piece limiting the collection angle, we collect scattered electrons an order of magnitude greater (and in area two orders greater) than in TEMs (see Supplementary Information ``Angular Range").

Collecting high angle electrons scattering is beneficial. It is used for phase identification, Euler angle measurements, and orientation mapping \cite{krieger_lassen_relative_1996,ram_phase_2018}. Unlike single scattered electrons, which form Bragg spots, multiple scattered electrons (diffusely and coherently scattered) form Kossel cones. Kossel cones intersect with Ewald sphere to produce Kikuchi bands. Kikuchi bands are great circles on the diffraction sphere and observed as projected lines on the detector. They intersect at zones and collectively serve as a map for identifying the crystal phase and orientation. 

Having a large collection angle (typically more than three poles are visible), i.e. having a wide reciprocal space field of view, is also desirable for detecting small crystal structure changes. The width of a Kikuchi band equals twice the Bragg angle $\theta_\text{B}$, which is sensitive to hydrostatic strain. The angle between two Kikuchi bands equals to the corresponding interplanar angle, which is sensitive to deviatoric strain. Temperature and stress changes induce both types. A fitting procedure is developed to fit a Kikuchi pattern collectively, which yields better precision than measuring a single Kikuchi line \cite{wu_novel_2012,gnabasik_fast_2025}. Fitting the full Kikuchi pattern takes advantage of all available information, which enables precise temperature and strain measurements beyond what previously has only been possible with TEMs.

\subsection*{Experiment}
\begin{figure}
	\begin{center}
		\includegraphics[width=\textwidth]{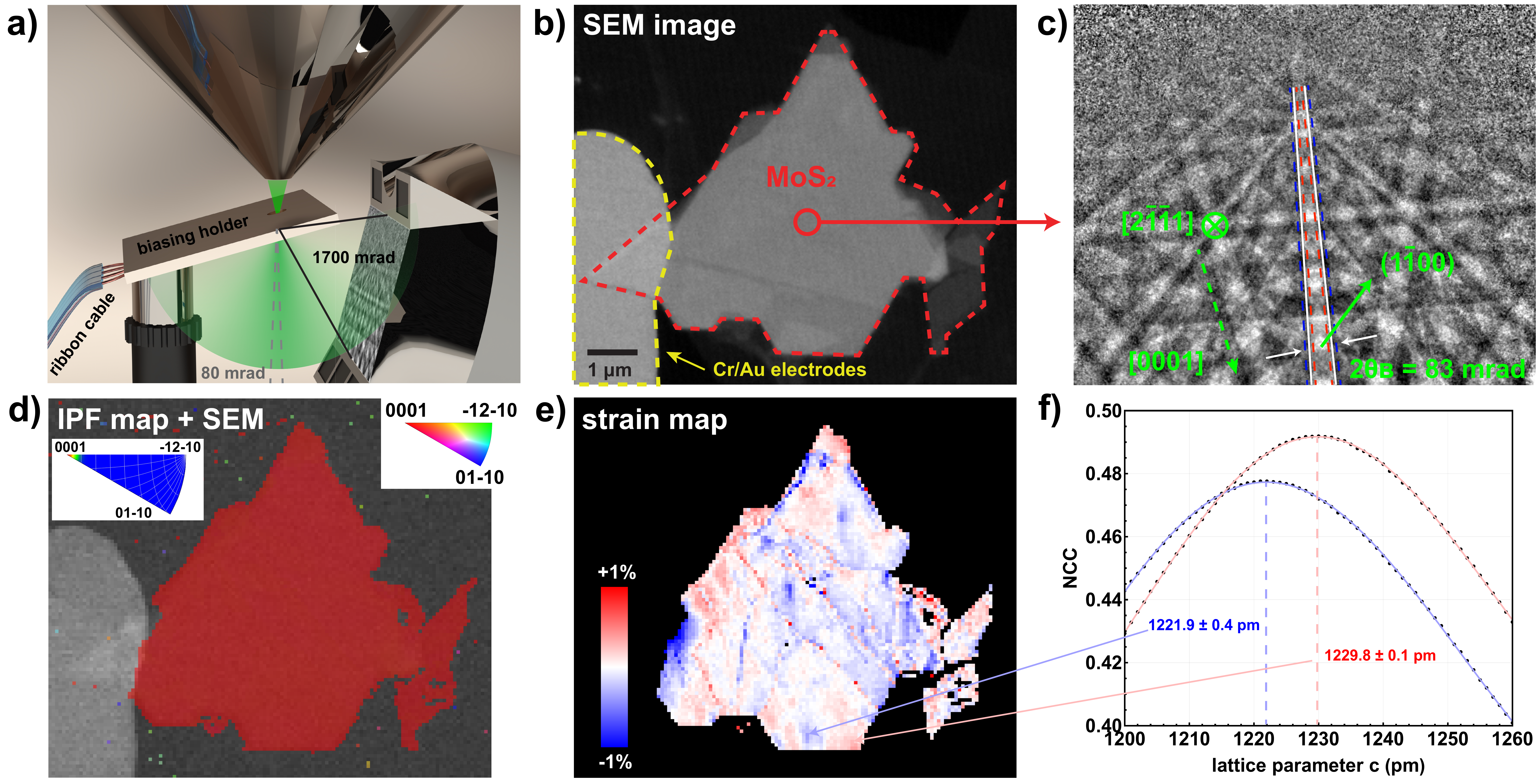}
		\caption{\textbf{Strain mapping with TKD.} a) Diagram of \textit{in-situ} biasing setup in the SEM chamber. The sample is attached to the back of the holder, and the electrical connection is established through a ribbon cable (see Supplementary Information ``Microchip Design"). The setup utilizes an off-axis TKD detector geometry. The black solid lines and gray dashed lines indicate collection angles for typical TKD and 4D-STEM systems, respectively. b) SEM image of a MoS$_2$ flake collected by forward-scatter detectors. c) A TKD pattern from the middle of the MoS$_2$ flake. A high-symmetry zone axis [0001] is at the bottom of the pattern outside of the collection region. The band highlighted with solid white lines corresponds to the family of lattice plane $(1\bar{1}00)$. The width of the Kikuchi bands varies with strain, with red and blue dashed lines indicating the directions of change that correspond to positive and negative strain respectively. d) Inverse pole figure (IPF) map overlayed with SEM image and IPF (inset) showing (0001) alignment. e) Strain map of the MoS$_2$ flake. f) Normalized cross-correlation coefficient (NCC) plot from two different pixels in \textit{e)}, the position of the max NCC is the best fit lattice parameter.}
		\label{fig:intro}
	\end{center}
\end{figure}

To demonstrate our capability of TKD strain and temperature mapping, we develop a \textit{in-situ} biasing setup for SEM (Helios G4 PFIB UXe DualBeam FIB/SEM). A custom holder with electrical contacts is tilted at \SI{20}{\degree}, and an electron-transparent sample is mounted upside down on the back of the holder (Fig.\ \ref{fig:intro}a). This geometry allows the sample (instead of the support membrane) to be the exit surface, improving the quality of the collected Kikuchi patterns \cite{rice_specimen-thickness_2014}. An Oxford Symmetry EBSD detector collects TKD patterns at a detector distance of 14.4 mm. This off-axis geometry gives an equivalent collection semi-angle of 740 mrad and covers a solid angle two orders of magnitude larger than achievable in a TEM. 

We fabricate the electron-transparent biasing chip from a \SI{200}{\text{-}\micro\metre}-thick silicon wafer with a 20-nm-thick silicon nitride (Si$_3$N$_4$) membrane on each side. Electron-transparent windows are defined from the back side of the wafer with optical lithography, and the Si$_3$N$_4$ in the corresponding region is removed by reactive ion etching (RIE). Si is removed from the back using KOH wet etch, leaving the 20-nm-thick Si$_3$N$_4$ on the top side that serves as the support membrane. Cr/Au electrodes are patterned on the front side with optical lithography and deposited with electron-beam evaporation. 

A molybdenum disulfide (MoS$_2$) flake is dry-transferred onto the electron-transparent window (Fig.\ \ref{fig:intro}b). Electrons with 20~keV kinetic energy are focused to form a probe with 4.3~mrad convergence semi-angle. As the electron probe scans the sample with a step size of 72~nm and a dwell time of 49~ms, a Kikuchi pattern is collected at every beam position (Fig.\ \ref{fig:intro}c). The MoS$_2$ [0001] zone is below the bottom of the image (outside the FOV), and the $(1\bar{1}00)$ band is highlighted in white. The acquired TKD map is first indexed with Oxford AZtecHKL using the Hough indexing algorithm and then refined with Oxford MapSweeper using cross-correlation-based pattern matching \cite{trimby_characterisation_2024}. The Hough indexing algorithm returns a uniform (0001) orientation across the MoS$_2$ flake as expected.

We use cross-correlation-based pattern matching to further refine the local orientation (Fig.\ \ref{fig:intro}d). A master pattern, which is the diffraction sphere equal-area projected onto 2D Cartesian coordinates, is generated for a crystal structure. A cross-correlation algorithm compares each experimental TKD pattern to simulated patterns with different orientations subsampled from the master pattern. The normalized cross-correlation coefficient (NCC) ranges from -1 to +1 and quantifies the similarity between experimental and simulated patterns. It is maximized to find the best match orientation \cite{chen_dictionary_2015,friedrich_application_2018,de_graef_dictionary_2020}. Cross-correlation-based pattern matching yields high-precision orientation measurements at the cost of higher computation power than the traditional Hough indexing method. The two-step analysis routine we use takes the phase and orientation determined by the Hough indexing, which is close to the optimal solution, as the starting point for the cross-correlation-based optimization. Providing a starting point close to the best match solution dramatically reduced the number of iterations to converge to a solution, therefore, combines the speed of the Hough indexing and the precision of cross-correlation analysis. This reduction in the necessary computation power allows us to expand the parameter space beyond the crystal orientation. 

In Kikuchi diffraction thermometry, we extend the parameter space of the pattern matching routine to include lattice constant variations by generating a set of master patterns with different lattice constants. For each experimental TKD pattern, the orientation is optimized for every master pattern in the set, and the NCC of the best matching orientation for each master pattern is plotted as a function of the corresponding lattice constant (Fig.\ \ref{fig:intro}f). We fit the NCC curve to a Gaussian function with a linear background and take the position of the maximum to be the local lattice constant at the corresponding beam position. The local variation in the lattice constant c is determined with sub-pm precision and is converted to strain (Fig.\ \ref{fig:intro}e) to reveal deformation invisible in the orientation map.

\begin{figure}
	\begin{center}
		\includegraphics[width=\textwidth]{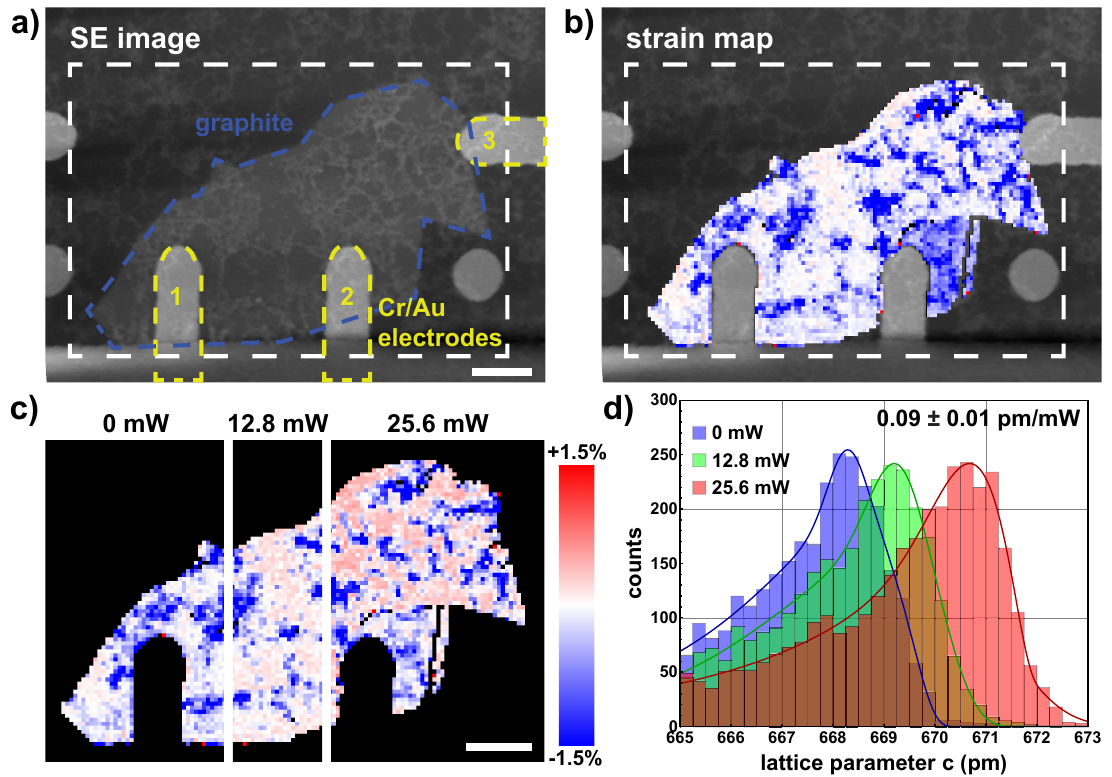}
		\caption{\textbf{Detection of thermal expansion.} a) SEM image of a graphite flake connected to three Cr/Au electrodes. Electrode 1 and 2 are used to bias the graphite flake. The white box highlights the region where TKD maps are acquired. b) Strain map of the graphite flake without electrical bias, overlaid on the SEM image. c) Strain maps of the graphite flake at different applied heating powers (see Supplementary Information ``Strain Maps" for complete maps). The strain increases with applied power due to thermal expansion. d) Histogram of lattice parameter $c$ measured at different heating powers. All scale bars are \SI{5}{\micro\metre}. The color bar applies to both \textit{b)} and \textit{c)}.}
		\label{fig:const_power}
	\end{center}
\end{figure}

To demonstrate the temperature sensitivity in TKD, we prepare a graphite flake connected to two Cr/Au electrodes on a Si$_3$N$_4$ electron-transparent window (Fig.\ \ref{fig:const_power}a). A current from a Keithley 6430 source measure unit provides the power Joule heating the sample. TKD maps are acquired at three different powers, 0 mW, 12.8 mW, and 25.6 mW. Applying the cross-correlation pipeline to the three TKD maps yields a room temperature baseline and two maps at elevated temperatures (Fig.\ \ref{fig:const_power}b). Comparison of the three strain maps (Fig.\ \ref{fig:const_power}c) reveals a temperature-induced positive strain of $0.5\%$ due to thermal expansion. The histogram of the graphite lattice parameter $c$ shows a significant right-hand shift as the heating power increases. On average, for every mW power sourced through the device, the lattice parameter $c$ expands $0.094\pm0.009$~pm, which corresponds to $0.0140\pm0.0013$\%/mW temperature-induced strain. Considering the 36.5~GPa out-of-plane Young's modulus of graphite \cite{blakslee_elastic_1970}, we determine the thermal stress to be $5.1\pm0.5$~MPa/mW. 

\begin{figure}
	\begin{center}
		\includegraphics[width=0.8\textwidth]{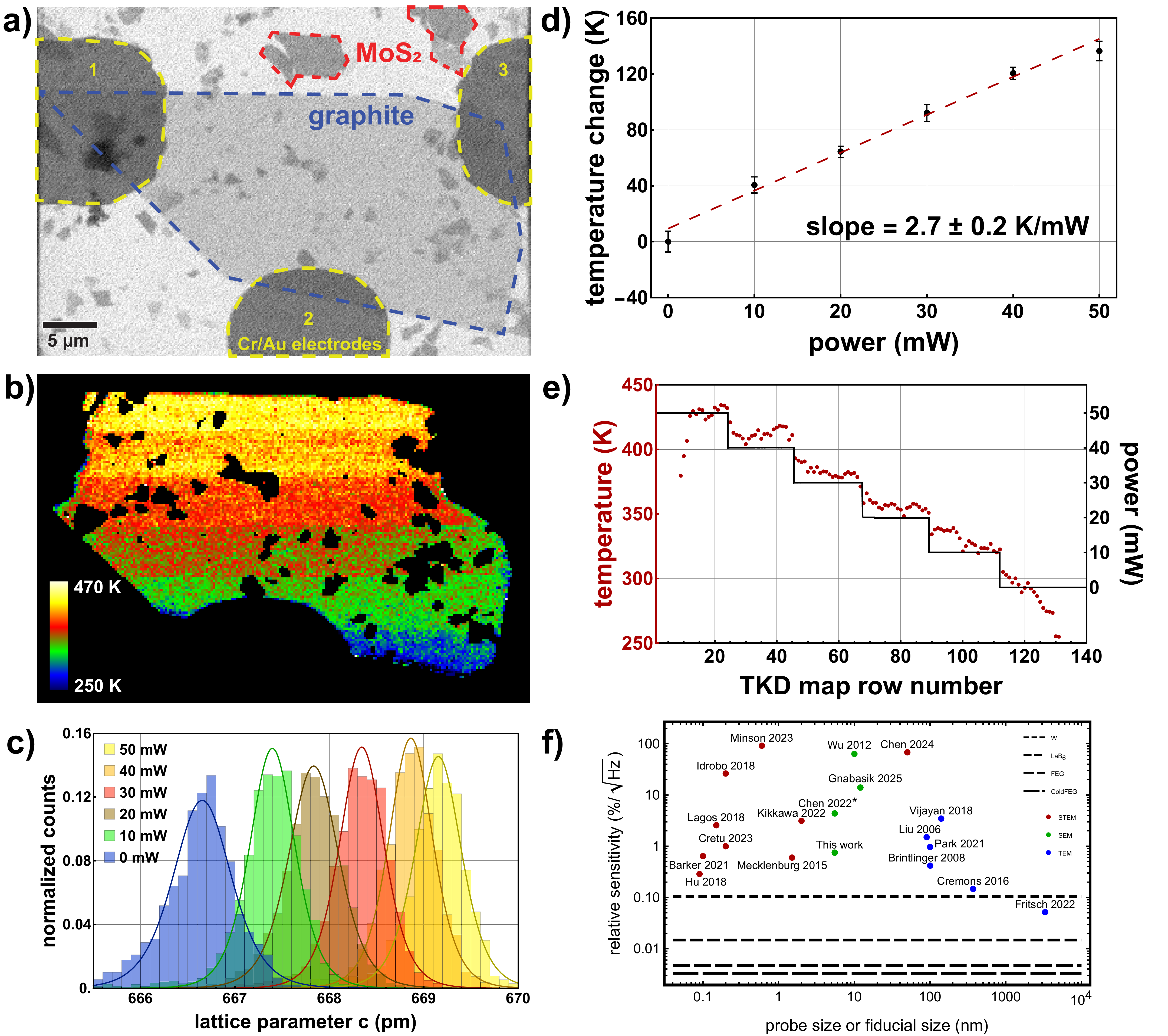}
		\caption{\textbf{\textit{In-situ} temperature mapping while stepping the power.} a) A graphite flake connected to three Cr/Au electrodes. Electrodes 1 and 2 are used to bias the graphite flake. MoS$_2$ flakes scattered around the window are ignored in this study. b) Temperature map of the graphite flake while the heating power is varied six times during a TKD scan. The six different temperatures are visually distinguishable. c) Histogram of the measured lattice parameter $c$ for the six different regions in \textit{b)}. The peaks correspond to different heating powers are clearly separated. d) Temperature change as a function of biasing power. A linear relation is observed. e) Row-average temperature plotted with corresponding heating power. f) A summary of temperature measurements done in electron microscopes \cite{liu_novel_2006,vijayan_situ_2018,park_direct_2021,brintlinger_electron_2008,cremons_direct_2016,fritsch_sub-kelvin_2022,wu_novel_2012,wu_development_2013,gnabasik_fast_2025,chen_detecting_2022,hu_mapping_2018,barker_automated_2021,mecklenburg_nanoscale_2015,cretu_nanometer-level_2023,lagos_thermometry_2018,kikkawa_optical_2022,idrobo_temperature_2018,minson_quantitative_2023,chen_estimation_2024}. The probe size or fiducial size limits the spatial resolution. However, the spatial resolution for temperature mapping is also limited by the phonon mean free path \cite{mecklenburg_nanoscale_2015}, which is typically a few nanometers or above. Therefore, a probe size below 1~nm does not necessarily mean an improved spatial resolution for temperature mapping. The dashed lines represent relative sensitivity limit for different types of electron guns (see Supplementary Information ``Gun Brightness Limited Measurement Sensitivity" section for details). }
		\label{fig:ramp}
	\end{center}
\end{figure}

We also vary the power within a single acquisition, this time with a different sample that does not exhibit significant strain ($\sim$0.1\%) at room temperature. We source 50~mW power through graphite from Cr/Au electrodes 1 and 2 (Fig.\ \ref{fig:ramp}a) prior to the start of the TKD acquisition. During the acquisition, the power is stepped down by 1~ mW every two to three minutes until 0~mW is reached. Determining the out-of-plane lattice parameter with the cross-correlation routine, we map the temperature-induced deformation throughout the acquisition and, knowing the coefficient of thermal expansion (CTE, $\alpha_c\approx$ 25~ppm/~$^{\circ}$C) \cite{pierson_handbook_1993,morgan_thermal_1972,zhao_review_2022}, convert it to a temperature map (Fig.\ \ref{fig:ramp}b). The temperature map is not masked, and MoS$_2$ flakes and Cr/Au electrodes are automatically excluded during phase identification. The map does not represent the spatial distribution of the temperature while the entire sample is in thermal equilibrium; instead, it shows the different temperatures of the sample while it is held at different heating powers throughout the TKD acquisition. Six horizontal stripes are seen parallel to the SEM’s fast scan direction, and the row averaged profile (Fig.\ \ref{fig:ramp}e) shows the correspondence between temperature and power. The temperature measured within each stripe is uniform, but a lower temperature is recorded around the edges of the graphite flake due to the proximity to heat sinks (i.e. Cr/Au electrodes) or a negative strain of approximately $0.1$\% due to edge warping. Together with an uncorrected internal strain $\sim$0.1\% at room temperature, the effects mentioned above are responsible for the roll-off at the beginning and the end of the row averaged profile (Fig.\ \ref{fig:ramp}e). The out-of-plane lattice parameter shifts linearly as a function of power (Fig.\ \ref{fig:ramp}c), and the average lattice parameter in each region is determined by a Gaussian fit. Taking the 0~mW region as a reference and calculating the corresponding temperature change with CTE, we determine that each mW increases the temperature by $2.7\pm0.2$~K (Fig.\ \ref{fig:ramp}d). 

Temperature sensitivity and spatial resolution are two critical parameters for evaluating the performance of a thermometry technique. For the dataset shown in Fig.\ \ref{fig:ramp}, we have a temperature uncertainty of $\delta T = 14$ K for each acquired TKD pattern and a 24-ms exposure time, which gives a temperature sensitivity of 2.2~K$/\sqrt{\text{Hz}}$ (or a relative sensitivity of 0.75~\%$/\sqrt{\text{Hz}}$, see Supplementary Information ``Gun Brightness Limited Measurement Sensitivity" section for definition), on par with the sensitivity achieved with TEMs (Fig.\ \ref{fig:ramp}f). The probe size allows us to compare the theoretical limits of spatial resolution of a variety of techniques independent of the sample. The Fig.\ \ref{fig:ramp} dataset is acquired with an aperture of \SI{64}{\micro\metre} and a beam energy of 15~keV, which yields a probe size of 5.5~nm (see Supplementary Information ``Probe Size" section). This 5.5~nm resolution limit is superior to that achieved using TEM mode with a parallel beam or fiducial markers \cite{liu_novel_2006,vijayan_situ_2018,park_direct_2021,brintlinger_electron_2008,cremons_direct_2016,fritsch_sub-kelvin_2022}, but it is eclipsed by the resolution limit of STEM-based thermometries \cite{hu_mapping_2018,barker_automated_2021,mecklenburg_nanoscale_2015,cretu_nanometer-level_2023,lagos_thermometry_2018,kikkawa_optical_2022,idrobo_temperature_2018,minson_quantitative_2023,chen_estimation_2024}. However, the actual spatial resolution also depends on the phonon mean free path and the interaction volume, which are related to the material under investigation, the sample thickness and the electron beam energy. The phonon mean free path in different materials typically ranges from a few nanometers to a few microns; therefore, the sub-nanometer probe size achievable in STEMs only brings none-to-marginal improvement in spatial resolution for temperature mapping. Among SEM-based techniques, EBSD generally has an interaction volume of 0.1~–~\SI{1}{\micro\metre}, which largely limits its spatial resolution. TKD can achieve an interaction volume less than 10~nm by collecting the transmitted electrons after a thin sample ($<$100~nm). The interaction volume of our setup is estimated to be 5-16~nm (see Supplementary Information ``Interaction Volume" section). 


\begin{figure}
	\begin{center}
		\includegraphics[width=\textwidth]{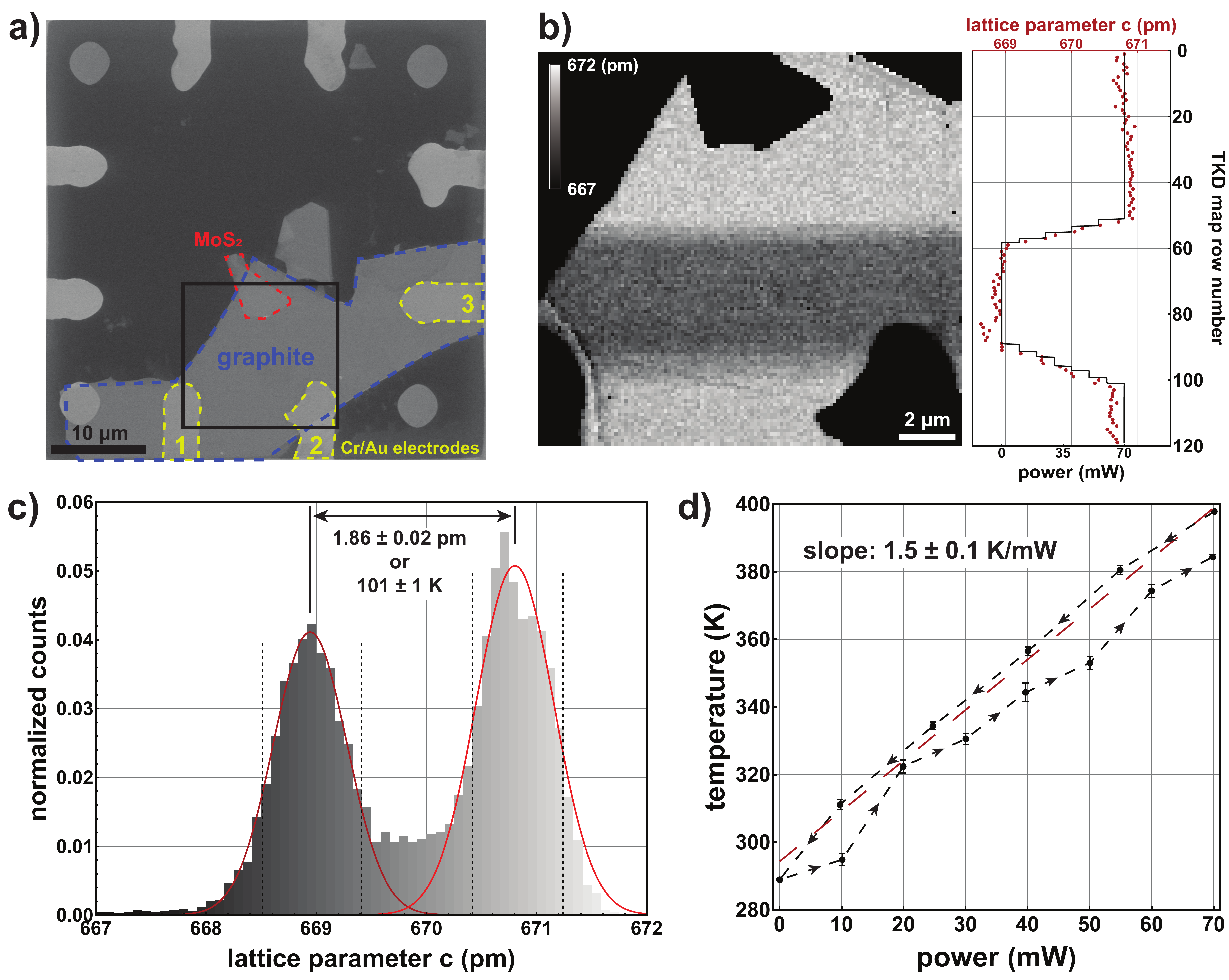}
		\caption{\textbf{Continuous temperature ramp.} a) SEM image of a graphite flake in contact with three Cr/Au electrodes. The black rectangle indicates the field of view where TKD mapping is performed. b) A map of lattice parameter $c$ of the graphite flake derived from the TKD map as the heating power varies. The row-average of the lattice parameter tracks the varying heating power, indicating the temperature induced thermal expansion. c) Histogram of \textit{b)}. The two peaks, corresponding to 0 mW and 70 mW, respectively, are separated by $1.86\pm0.02$~pm, which corresponds to a temperature difference of $101\pm1$~K. The dashed lines indicate the fit window around each peak, and the red curves represent fits to a Gaussian distribution. d) Temperature versus heating power. The temperature change induced by Joule heating is $1.5\pm0.1$ K/mW. Black arrows indicate the chronology of the power ramp.}
		\label{fig:stripe}
	\end{center}
\end{figure}

To demonstrate that we can track a near-continuous temperature ramp, we prepared another exfoliated graphite sample connected to three Cr/Au electrodes (Fig.\ \ref{fig:stripe}a). A bias is applied across electrodes 1 and 2 to heat up the sample while a TKD map is acquired with a 20~kV electron beam. 70~mW power is supplied to the graphite flake at the beginning of the acquisition and is maintained for three minutes before it is decreased to 0~mW. The sample is held at room temperature for 2 minutes before ramping the power back to 70~mW. The microscopic system has a small thermal mass, and the lattice parameter instantly reflects temperature changes. After feeding the acquired TKD map to the cross-correlation pipeline for lattice parameter determination, a correlation between interlayer spacing and the power sourced through the sample is revealed (Fig.\ \ref{fig:stripe}b). The row-averaged line profile is sensitive to sub-pm changes in lattice parameters, which corresponds to a strain sensitivity on the $0.01\%$ level, on par with the precision achieved with 4D STEM strain analysis \cite{mahr_accurate_2021}. The histogram of the lattice parameter map shows two well-resolved peaks corresponding to the regions acquired at 0~mW and 70~mW, respectively (Fig.\ \ref{fig:stripe}c). The changes in the lattice parameter and temperature are measured with 0.02~pm and 1~K precision, respectively. The measured temperature is linear in the heating power (Fig.\ \ref{fig:stripe}d), as expected from Fourier's law of heat conduction. Three data points at 0~mW and 70~mW come from the large bright, dark, bright regions, and all other data points in between come from the two transitional regions where the power is varying. The ramp-down branch in Fig.\ \ref{fig:stripe}d shows great linearity, while the ramp-up branch is slightly disturbed by temperature-independent strain near the Cr/Au electrodes. A small temperature difference of about 10~K is observed between the two 70~mW regions, which we attribute to a spatial temperature gradient across the graphite flake. The bottom half of the field of view is closer to the edge of the electron-transparent window. Outside the electron-transparent window, the sample thickness quickly increases to \SI{200}{\micro\metre} due to the silicon substrate, which serves as a heat sink at room temperature. Considering the sample geometry and the in-plane thermal conductivity of graphite, the heat flow corresponding to a 10~K temperature difference is roughly 1~mW, in line with our expectation. The temperature sensitivity achieved with this dataset is 2.5~K$/\sqrt{\text{Hz}}$, matching the result of the dataset in Fig.\ \ref{fig:ramp}.


\subsection*{Results}

Precise strain measurement using EBSD was first proposed by Wilkinson, Meaden, and Dingley in 2006 \cite{wilkinson_high-resolution_2006}. This technique (later on referred to as high-resolution EBSD or HR-EBSD) measures the shift of zone axes in an experimental pattern with respect to a reference pattern and solves for the strain components that induce such a shift. HR-EBSD has been demonstrated to have $\sim$0.01\% precision and is sensitive to changes in c/a ratio \cite{britton_high_2012,speller_microstructural_2012}. Fundamentally, shifts in the zone axes are induced by changes in interplanar angles. Because HR-EBSD only captures changes in interplanar angles, it is insensitive to hydrostatic strain. Moreover, the orientation optimization and strain measurement of HR-EBSD are performed as two separate steps, which means that simultaneous optimization in lattice rotation and strain is not guaranteed. Kikuchi diffraction thermometry searches for the best match solution in a parameter space of both the crystal orientation and the lattice parameter, improving the accuracy of both lattice rotation and strain determinations while achieving $\sim$0.01\% precision in strain measurements. KDTh also takes into account all information in the Kikuchi pattern during optimization; therefore, it is sensitive to changes in the width of the Kikuchi bands induced by hydrostatic strain. A similar idea has recently been proposed to detect hydrostatic strain in the HR-EBSD framework, where a library of master patterns with different accelerating voltages (instead of different lattice constants) is generated and a hydrostatic strain measurement with $\sim$0.1\% precision is demonstrated \cite{shi_measuring_2025}.

The work presented is the first to show SEM-based temperature mapping. Compared to other SEM-based thermometries, a higher temperature sensitivity is achieved with KDTh (2.2~K/$\sqrt{\text{Hz}}$) than Chen's 21~K/$\sqrt{\text{Hz}}$ \cite{chen_detecting_2022}, Wu's 170~K/$\sqrt{\text{Hz}}$ \cite{wu_novel_2012,wu_development_2013}, and Gnabasik's 41~K/$\sqrt{\text{Hz}}$ \cite{gnabasik_fast_2025} (more details in Fig.\ \ref{fig:ramp}f). The difference is caused by the choice of platform (EBSD vs.\ TKD), sample composition (crystal quality), technique (how strain is measured), and camera architecture. The platform and sample are chosen according to the need (electron-transparent samples are required for TKD). The choice is between technique and camera architecture. Wu and Gnabasik's works \cite{wu_novel_2012,gnabasik_fast_2025} are based on the Debye-Waller effect and TDS. Although temperature measurement with TDS can be more general (as crystallinity is not required \cite{libera_temperature-dependent_1996}), all coherent scattering information that is temperature dependent and has a high signal-to-noise ratio is left out. Chen's 2022 work \cite{chen_detecting_2022} uses Hough indexing-based MAD optimization instead of cross-correlation-based NCC optimization, which has one order of magnitude less angular precision (see Supplementary Information ``Kikuchi Diffraction Thermometry with Hough MAD Optimization" section for a direct comparison) and is insensitive to changes in Kikuchi band widths. However, Hough indexing-based MAD optimization requires significantly less computational power. On contemporary TKD/EBSD systems where the Hough indexing speed of thousands of patterns per second is routinely achieved, the Hough indexing-based MAD optimization can be implemented for real-time analysis.

The technique used is detailed above, so a short discussion of cameras can provide a final piece of insight to further precision. Older EBSD/TKD systems use lens-coupled CCD cameras, which have now given way to fiber-couple CCDs. These systems offer decent read noise performance with large dynamic range (16-bit), but are slow compared to two other camera types. Direct electron detectors in counting mode have a low dynamic range (and in integrating mode lower than CCDs), but they are fast and very sensitive. A recent study also shows that electrons that undergo high energy loss yield lower band contrast, so energy filtering is necessary to maintain the quality of Kikuchi patterns in counting mode \cite{ventura_energy-resolved_2025}. As EBSD/TKD often requires large currents, direct electron detectors are not always the optimal solution. Hybrid-pixel detectors have both a large dynamic range, high speed, and low read noise. They are potentially the best solution, in particular at lower kVs, but have not been widely implemented yet. Future work would see how much these latter types of detectors improve precision in these measurements.

\subsection*{Discussion}

By expanding the parameter space of the cross-correlation-based TKD indexing pipeline, we detect temperature changes on micrometer-scale graphite devices in an SEM with Kikuchi diffraction. We achieved a temperature sensitivity of 2.2~K$/\sqrt{\text{Hz}}$ (or 0.75~\%$\sqrt{\text{Hz}}$), on par with that achieved in TEMs \cite{park_direct_2021,brintlinger_electron_2008,cremons_direct_2016,fritsch_sub-kelvin_2022,hu_mapping_2018,barker_automated_2021,mecklenburg_nanoscale_2015,cretu_nanometer-level_2023}. Kikuchi diffraction thermometry demonstrated in this work can be paired with either SEM TKD or EBSD (with resolution trade-off) to offer temperature mapping or volumetric strain mapping capability in parallel with the obtaining of crystalline phase and orientation information. By probing the lattice spacing through Kikuchi diffraction, the spatial resolution of KDTh is limited by the size of the electron probe and the interaction volume. The 5-16~nm interaction volume (see Supplementary Information ``Probe Size" and ``Interaction Volume" sections) used in this study is close to the smallest possible feature size for temperature mapping in common materials, which is limited by the mean free path of phonons \cite{mecklenburg_nanoscale_2015}. The capability to measure temperature at the nanometer scale using far-field SEM optics is important because not all samples are suitable for TEM, and a wider field of view and a lower cost are also beneficial.

Kikuchi diffraction thermometry has high potential to extend to polycrystalline samples. Thanks to existing pipelines for phase and orientation mapping using Kikuchi patterns, KDTh is easier to work with polycrystalline samples than 4D STEM, where indexing the patterns reliably remains a major challenge. However, characterizing polycrystalline samples with extremely small grains ($<$10~nm) might remain the regime of TEM-based techniques. Limited by SEM optics and interaction volume, TKD and EBSD can only easily resolve grains down to 10~nm and several hundreds of nanometers, respectively \cite{lee_transmission_2017,zhu_-depth_2018}. Moreover, KDTh provides a general methodology for adapting commercially available TKD/EBSD analysis pipelines for more specific research or applications. By customizing the library of simulated patterns for pattern matching, structural information such as axial strain, shear strain, film thickness, or even physical properties such as the Debye-Waller factor can potentially be extracted. The implications for imaging temperature in microelectronic devices without their destruction opens up many avenues of further thermal investigation into thermal transport and heat dissipation.


\clearpage 

%
\bibliography{20250528_TKD} 

\begin{thebibliography}{10}
\providecommand{\url}[1]{\texttt{#1}}
\expandafter\ifx\csname urlstyle\endcsname\relax
  \providecommand{\doi}[1]{doi:\discretionary{}{}{}#1}\else
  \providecommand{\doi}{doi:\discretionary{}{}{}\begingroup
  \urlstyle{rm}\Url}\fi

\bibitem{stopka_surface_1995}
M.~Stopka, L.~Hadjiiski, E.~Oesterschulze, R.~Kassing, Surface investigations
  by scanning thermal microscopy. \emph{Journal of Vacuum Science \& Technology
  B: Microelectronics and Nanometer Structures Processing, Measurement, and
  Phenomena} \textbf{13}~(6), 2153--2156 (1995).

\bibitem{menges_temperature_2016}
F.~Menges, \emph{et~al.}, Temperature mapping of operating nanoscale devices by
  scanning probe thermometry. \emph{Nature Communications} \textbf{7}~(1),
  10874 (2016).

\bibitem{presotto_super-resolution_2024}
L.~Presotto, \emph{et~al.}, Super-{Resolution} {Photothermal} {Imaging} at the
  {Microscale} by {Model}-{Based} {Image} {Reconstruction}. \emph{Advanced
  Intelligent Systems} \textbf{6}~(1), 2300510 (2024).

\bibitem{reparaz_novel_2014}
J.~S. Reparaz, \emph{et~al.}, A novel contactless technique for thermal field
  mapping and thermal conductivity determination: {Two}-{Laser} {Raman}
  {Thermometry}. \emph{Review of Scientific Instruments} \textbf{85}~(3),
  034901 (2014).

\bibitem{kim_examination_2003}
H.~J. Kim, K.~D. Kihm, J.~S. Allen, Examination of ratiometric laser induced
  fluorescence thermometry for microscale spatial measurement resolution.
  \emph{International Journal of Heat and Mass Transfer} \textbf{46}~(21),
  3967--3974 (2003).

\bibitem{cremons_direct_2016}
D.~R. Cremons, D.~J. Flannigan, Direct in situ thermometry: {Variations} in
  reciprocal-lattice vectors and challenges with the {Debye}–{Waller} effect.
  \emph{Ultramicroscopy} \textbf{161}, 10--16 (2016).

\bibitem{fritsch_sub-kelvin_2022}
B.~Fritsch, \emph{et~al.}, Sub-{Kelvin} thermometry for evaluating the local
  temperature stability within in situ {TEM} gas cells. \emph{Ultramicroscopy}
  \textbf{235}, 113494 (2022).

\bibitem{mecklenburg_nanoscale_2015}
M.~Mecklenburg, \emph{et~al.}, Nanoscale temperature mapping in operating
  microelectronic devices. \emph{Science} \textbf{347}~(6222), 629--632 (2015).

\bibitem{mecklenburg_visualizing_2021}
M.~Mecklenburg, B.~T. Zutter, X.~Y. Ling, W.~A. Hubbard, B.~C. Regan,
  Visualizing the {Electron} {Wind} {Force} in the {Elastic} {Regime}.
  \emph{Nano Letters} \textbf{21}~(24), 10172--10177 (2021).

\bibitem{wu_novel_2012}
X.~Wu, R.~Hull, A novel nano-scale non-contact temperature measurement
  technique for crystalline materials. \emph{Nanotechnology} \textbf{23}~(46),
  465707 (2012).

\bibitem{gnabasik_fast_2025}
R.~Gnabasik, \emph{et~al.}, Fast nanothermometry based on direct electron
  detection of electron backscattering diffraction patterns (2025).

\bibitem{shen_spatial_2019}
Y.~Shen, \emph{et~al.}, Spatial {Resolutions} of {On}-{Axis} and {Off}-{Axis}
  {Transmission} {Kikuchi} {Diffraction} {Methods}. \emph{Applied Sciences}
  \textbf{9}~(21), 4478 (2019).

\bibitem{zhang_comparison_2025}
T.~Zhang, L.~Berners, J.~Holzer, T.~B. Britton, Comparison of {Kikuchi}
  diffraction geometries in the scanning electron microscope. \emph{Materials
  Characterization} \textbf{222}, 114853 (2025).

\bibitem{shevitski_dark-field_2013}
B.~Shevitski, \emph{et~al.}, Dark-field transmission electron microscopy and
  the {Debye}-{Waller} factor of graphene. \emph{Physical Review B}
  \textbf{87}~(4), 045417 (2013).

\bibitem{khidirov_root-mean-square_2021}
I.~Khidirov, S.~D. Rakhmanov, S.~A. Makhmudov, Root-{Mean}-{Square} {Amplitude}
  of {Zero}-{Point} {Vibrations} in a {Crystal} {\textbar} {Russian} {Physics}
  {Journal} (2021).

\bibitem{tsuno_design_1998}
K.~Tsuno, D.~A. Jefferson, Design of an objective lens pole piece for a
  transmission electron microscope with a resolution less than 0.1nm at 200kV.
  \emph{Ultramicroscopy} \textbf{72}~(1), 31--39 (1998).

\bibitem{liu_depth_2019}
J.~Liu, S.~Lozano-Perez, A.~J. Wilkinson, C.~R.~M. Grovenor, On the depth
  resolution of transmission {Kikuchi} diffraction ({TKD}) analysis.
  \emph{Ultramicroscopy} \textbf{205}, 5--12 (2019).

\bibitem{krieger_lassen_relative_1996}
N.~C. Krieger~Lassen, The relative precision of crystal orientations measured
  from electron backscattering patterns. \emph{Journal of Microscopy}
  \textbf{181}~(1), 72--81 (1996).

\bibitem{ram_phase_2018}
F.~Ram, M.~De~Graef, Phase differentiation by electron backscatter diffraction
  using the dictionary indexing approach. \emph{Acta Materialia} \textbf{144},
  352--364 (2018).

\bibitem{rice_specimen-thickness_2014}
K.~Rice, R.~Keller, M.~Stoykovich, Specimen-thickness effects on transmission
  {Kikuchi} patterns in the scanning electron microscope. \emph{Journal of
  Microscopy} \textbf{254}~(3), 129--136 (2014).

\bibitem{trimby_characterisation_2024}
P.~Trimby, \emph{et~al.}, The characterisation of dental enamel using
  transmission {Kikuchi} diffraction in the scanning electron microscope
  combined with dynamic template matching. \emph{Ultramicroscopy} \textbf{260},
  113940 (2024).

\bibitem{chen_dictionary_2015}
Y.~H. Chen, \emph{et~al.}, A {Dictionary} {Approach} to {Electron}
  {Backscatter} {Diffraction} {Indexing}. \emph{Microscopy and Microanalysis}
  \textbf{21}~(3), 739--752 (2015).

\bibitem{friedrich_application_2018}
T.~Friedrich, A.~Bochmann, J.~Dinger, S.~Teichert, Application of the pattern
  matching approach for {EBSD} calibration and orientation mapping, utilising
  dynamical {EBSP} simulations. \emph{Ultramicroscopy} \textbf{184}, 44--51
  (2018).

\bibitem{de_graef_dictionary_2020}
M.~De~Graef, A dictionary indexing approach for {EBSD}. \emph{IOP Conference
  Series: Materials Science and Engineering} \textbf{891}~(1), 012009 (2020).

\bibitem{blakslee_elastic_1970}
O.~L. Blakslee, D.~G. Proctor, E.~J. Seldin, G.~B. Spence, T.~Weng, Elastic
  {Constants} of {Compression}‐{Annealed} {Pyrolytic} {Graphite}.
  \emph{Journal of Applied Physics} \textbf{41}~(8), 3373--3382 (1970).

\bibitem{pierson_handbook_1993}
H.~O. Pierson, \emph{Handbook of carbon, graphite, diamond, and fullerenes:
  properties, processing, and applications}, Materials science and process
  technology series (Noyes Publications, Park Ridge, N.J) (1993).

\bibitem{morgan_thermal_1972}
W.~C. Morgan, Thermal expansion coefficients of graphite crystals.
  \emph{Carbon} \textbf{10}~(1), 73--79 (1972).

\bibitem{zhao_review_2022}
L.~Zhao, J.~Tang, M.~Zhou, K.~Shen, A review of the coefficient of thermal
  expansion and thermal conductivity of graphite. \emph{New Carbon Materials}
  \textbf{37}~(3), 544--555 (2022).

\bibitem{liu_novel_2006}
Z.~Liu, Y.~Bando, J.~Hu, K.~Ratinac, S.~P. Ringer, A novel method for practical
  temperature measurement with carbon nanotube nanothermometers.
  \emph{Nanotechnology} \textbf{17}~(15), 3681--3684 (2006).

\bibitem{vijayan_situ_2018}
S.~Vijayan, In {Situ} {Investigation} of {Thermally} {Activated} {Processes}
  {Using} {MEMS}-{Based} {Devices}: {Practical} {Challenges} \& {Applications}.
  \emph{Doctoral Dissertations}  (2018).

\bibitem{park_direct_2021}
J.~Park, \emph{et~al.}, Direct {Quantification} of {Heat} {Generation} {Due} to
  {Inelastic} {Scattering} of {Electrons} {Using} a {Nanocalorimeter}.
  \emph{Advanced Science} \textbf{8}~(3), 2002876 (2021).

\bibitem{brintlinger_electron_2008}
T.~Brintlinger, Y.~Qi, K.~H. Baloch, D.~Goldhaber-Gordon, J.~Cumings, Electron
  {Thermal} {Microscopy}. \emph{Nano Letters} \textbf{8}~(2), 582--585 (2008).

\bibitem{wu_development_2013}
X.~Wu, \emph{Development of a scanning electron microscopy based temperature
  measurement technique}, Electronic thesis, Rensselaer Polytechnic Institute,
  Troy, NY (2013).

\bibitem{chen_detecting_2022}
Y.~Chen, J.~J. Lodico, X.~Y. Ling, B.~C. Regan, M.~Mecklenburg, Detecting
  {Temperature}-{Induced} {Strain} {Changes} using {In} {Situ} {Transmission}
  {Kikuchi} {Diffraction}. \emph{Microscopy and Microanalysis}
  \textbf{28}~(S1), 576--577 (2022).

\bibitem{hu_mapping_2018}
X.~Hu, \emph{et~al.}, Mapping {Thermal} {Expansion} {Coefficients} in
  {Freestanding} {2D} {Materials} at the {Nanometer} {Scale}. \emph{Physical
  Review Letters} \textbf{120}~(5), 055902 (2018).

\bibitem{barker_automated_2021}
A.~Barker, B.~Sapkota, J.~P. Oviedo, R.~Klie, Automated plasmon peak fitting
  derived temperature mapping in a scanning transmission electron microscope.
  \emph{AIP Advances} \textbf{11}~(3), 035330 (2021).

\bibitem{cretu_nanometer-level_2023}
O.~Cretu, \emph{et~al.}, Nanometer-level temperature mapping of {Joule}-heated
  carbon nanotubes by plasmon spectroscopy. \emph{Carbon} \textbf{201},
  1025--1029 (2023).

\bibitem{lagos_thermometry_2018}
M.~J. Lagos, P.~E. Batson, Thermometry with {Subnanometer} {Resolution} in the
  {Electron} {Microscope} {Using} the {Principle} of {Detailed} {Balancing}.
  \emph{Nano Letters} \textbf{18}~(7), 4556--4563 (2018).

\bibitem{kikkawa_optical_2022}
J.~Kikkawa, K.~Kimoto, Optical and acoustic phonon temperature measurements
  using electron nanoprobe and electron energy loss spectroscopy.
  \emph{Physical Review B} \textbf{106}~(19), 195431 (2022).

\bibitem{idrobo_temperature_2018}
J.~C. Idrobo, \emph{et~al.}, Temperature {Measurement} by a {Nanoscale}
  {Electron} {Probe} {Using} {Energy} {Gain} and {Loss} {Spectroscopy}.
  \emph{Physical Review Letters} \textbf{120}~(9), 095901 (2018).

\bibitem{minson_quantitative_2023}
P.~S. Minson, F.~Rivera, R.~Vanfleet, Quantitative {STEM}: {A} method for
  measuring temperature and thickness effects on thermal diffuse scattering
  using {STEM}/{EELS}, and for testing electron scattering models.
  \emph{Ultramicroscopy} \textbf{246}, 113684 (2023).

\bibitem{chen_estimation_2024}
Q.~Chen, \emph{et~al.}, Estimation of {Temperature} {Homogeneity} in
  {MEMS}-{Based} {Heating} {Nanochips} via {Quantitative} {HAADF}-{STEM}
  {Tomography}. \emph{Particle \& Particle Systems Characterization}
  \textbf{41}~(2), 2300070 (2024).

\bibitem{mahr_accurate_2021}
C.~Mahr, \emph{et~al.}, Accurate measurement of strain at interfaces in
  {4D}-{STEM}: {A} comparison of various methods. \emph{Ultramicroscopy}
  \textbf{221}, 113196 (2021).

\bibitem{wilkinson_high-resolution_2006}
A.~J. Wilkinson, G.~Meaden, D.~J. Dingley, High-resolution elastic strain
  measurement from electron backscatter diffraction patterns: {New} levels of
  sensitivity. \emph{Ultramicroscopy} \textbf{106}~(4), 307--313 (2006).

\bibitem{britton_high_2012}
T.~B. Britton, A.~J. Wilkinson, High resolution electron backscatter
  diffraction measurements of elastic strain variations in the presence of
  larger lattice rotations. \emph{Ultramicroscopy} \textbf{114}, 82--95 (2012).

\bibitem{speller_microstructural_2012}
S.~C. Speller, \emph{et~al.}, Microstructural analysis of phase separation in
  iron chalcogenide superconductors. \emph{Superconductor Science and
  Technology} \textbf{25}~(8), 084023 (2012).

\bibitem{shi_measuring_2025}
Q.~Shi, D.~Loisnard, M.~Mollens, S.~Roux, Measuring {Hydrostatic} {Dilation} by
  {High}-{Resolution} {Electron} {Back}-{Scatter} {Diffraction}.
  \emph{Microscopy and Microanalysis} \textbf{31}~(Supplement\_1), ozaf048.315
  (2025).

\bibitem{libera_temperature-dependent_1996}
M.~Libera, J.~A. Ott, K.~Siangchaew, Temperature-dependent high-angle electron
  scattering from a phase-separated amorphous {GeTe} thin film.
  \emph{Ultramicroscopy} \textbf{63}~(2), 81--91 (1996).

\bibitem{ventura_energy-resolved_2025}
N.~M.~D. Ventura, \emph{et~al.}, Energy-{Resolved} {EBSD} using a {Monolithic}
  {Direct} {Electron} {Detector} (2025).

\bibitem{lee_transmission_2017}
S.-Y. Lee, \emph{et~al.}, Transmission orientation imaging of copper thin films
  on polyimide substrates intended for flexible electronics. \emph{Scripta
  Materialia} \textbf{138}, 52--56 (2017).

\bibitem{zhu_-depth_2018}
X.~Zhu, \emph{et~al.}, In-depth evolution of chemical states and
  sub-10-nm-resolution crystal orientation mapping of nanograins in
  {Ti}(5~nm)/{Au}(20~nm)/{Cr}(3~nm) tri-layer thin films. \emph{Applied Surface
  Science} \textbf{453}, 365--372 (2018).

\end{thebibliography}
\bibliographystyle{sciencemag}

%
%
%
%
%
%


\section*{Acknowledgments}
This work was supported by the Semiconductor Research Corporation (SRC), National Science Foundation (NSF) award DMR-1611036, NSF STC award DMR-1548924 (STROBE), the BioPACIFIC MIP award DMR-1933487, and the Electron Imaging Center for Nanosystems (EICN) (RRID:SCR$\_$022900) at the University of California, Los Angeles’s California for NanoSystems Institute (CNSI). Data were collected at the Core Center of Excellence in Nano Imaging (University of Southern California). Samples were fabricated in the Integrated Systems Nanofabrication Cleanroom (ISNC) at the California NanoSystems Institute (CNSI).


\subsection*{Supplementary materials}
Collection Angles\\
Angular Range\\
Microchip Design\\
Strain Maps\\
Gun Brightness Limited Measurement Sensitivity\\
Probe Size\\
Interaction Volume\\
Kikuchi Diffraction Thermometry with Hough MAD Optimization\\
Figs. S1 to S6\\
Tables S1\\


\newpage


\renewcommand{\thefigure}{S\arabic{figure}}
\renewcommand{\thetable}{S\arabic{table}}
\renewcommand{\theequation}{S\arabic{equation}}
\renewcommand{\thepage}{S\arabic{page}}
\setcounter{figure}{0}
\setcounter{table}{0}
\setcounter{equation}{0}
\setcounter{page}{1} 

\end{document}